# Redy: Remote Dynamic Memory Cache [Extended Report]

Qizhen Zhang[*], Philip A. Bernstein[§], Daniel S. Berger[§], Badrish Chandramouli[§]
[*]University of Pennsylvania, [§]Microsoft Research
[*]qizhen@seas.upenn.edu, [§]{philbe, daberg, badrishc}@microsoft.com

## ABSTRACT

Redy is a cloud service that provides high performance caches using RDMA-accessible remote memory. An application can customize the performance of each cache with a service level objective (SLO) for latency and throughput. By using remote memory, it can leverage stranded memory and spot VM instances to reduce the cost of its caches and improve data center resource utilization. Redy automatically customizes the resource configuration for the given SLO, handles the dynamics of remote memory regions, and recovers from failures. The experimental evaluation shows that Redy can deliver its promised performance and robustness under remote memory dynamics in the cloud. We augment a production key-value store, FASTER, with a Redy cache. When the working set exceeds local memory, using Redy is significantly faster than spilling to SSDs.

## 1 INTRODUCTION
### 1.1 The Case for Remote Memory as Cache

Stateful cloud services store their states on secondary storage, such as server-local SSDs or a cloud storage service. Example storage services are database systems, key-value stores, and JSON stores. Stateful application services embed these data management systems, such as a directory service, document management system, or source code control system. To offer fast response time, these types of services store a subset of their states in memory caches.

When allocating a memory cache, a server need not be limited by its local available memory. It could use physical memory on other servers. Although remote memory has higher access time than the server's local memory due to network latency, there are many reasons why it can be an attractive choice.

First, a server's physical memory capacity is limited. It may have insufficient local memory available for a stateful service, particularly for its peak workloads. In this case, remote memory is the only option. Otherwise, its state has to be spilled to secondary storage, resulting in orders-of-magnitude performance degradation.

Second, some cloud services are satisfied if they can read records in a few microseconds (μs's), which does not require local memory performance. This is currently impossible to achieve with SSDs, but can be supported with fast data center networks [4].

Third, remote memory may be cheaper because it sits on lightly loaded servers. For example, Google, Facebook, and Alibaba report that as much as 50% of server memory in data centers is unutilized [25, 47]. An extreme case is *stranded memory*, which is unusable by its local server because its cores have all been allocated to local VMs. Stranded memory is essentially free. By using this otherwise wasted memory as a cache, a stateful service can run on smaller servers with less server-local cache, thereby reducing cost.

A fourth reason is the trend toward dedicated and disaggregated memory servers whose sole function is to offer memory to remote servers [22, 25, 36, 38, 47, 60]. This approach is becoming more feasible due to fast data center networks, whose point-to-point bandwidth is close to I/O bus bandwidth and is usually underutilized [45, 61]. Cloud service providers already disaggregate compute and storage. By disaggregating memory, they can fully benefit from this expensive resource. Most cloud vendors have not been forthcoming about their internal usage of this capability and do not yet support it for third-party users. However, Google recently reported that it uses disaggregated memory in its BigQuery service [38].

To be usable as a cache, remote memory must be accessible with very low latency. Remote direct memory access (RDMA) is the natural choice. RDMA is not as fast as local memory, but it is much faster than SSDs and requires little or no CPU involvement.

Today, the typical access time for main memory is 70 nanoseconds (ns) [57]. For RDMA it is a few μs [5, 7, 62]. For SSD it is ∼100 μs, but highly variable and often higher, due to garbage collection and concurrent writes. Although RDMA latency is 100x better than SSD, its bandwidth advantage is only 2x - 10x (e.g., SSDs are 16-24 Gbit/s and RDMA networks are 48-200 Gbit/s). Still, the difference is significant for applications that need high-throughput data access. Hence, RDMA-accessible remote memory is a natural choice for a cache sitting between these two layers of the memory hierarchy.

### 1.2 Contributions

There are two main challenges in using remote memory as a cache. The first is how tune RDMA configurations. The choice of optimal configuration depends on the application workload, processor and network characteristics, and service level objective (SLO). Misconfiguration can lead to poor performance. Tuning RDMA is known to be difficult. In a data center, it must be done dynamically, since the choice of processor and network distance between processors can vary. It is therefore important that this tuning be automated.

The second challenge is responding to changes in remote memory availability. A memory region might become unavailable because its server failed or because the memory region allocation was evictable and the system reclaimed it for local VMs. In both cases, the application that was using the cache must be dynamically reconfigured. It must operate without the cache or migrate the cache to another remote memory region and re-populate it.

We propose Redy, a new cloud cache service that efficiently utilizes stranded and unused server memory using RDMA. Unlike prior RDMA stores and caches, it handles failures and reclamations and allows users to customize cache performance. It also requires minimal changes to applications. Our contributions are as follows.

- Stranded memory analysis. We present the results of a study that shows stranded memory is significant and dynamic.

- An RDMA architecture for an SLO-based memory cache service. Unlike previous RDMA systems that optimize for specific performance targets, ours enables the user to customize the target. It automatically finds an RDMA configuration that satisfies the user-provided SLO and minimizes resource cost.
- Dynamic memory management. Redy is elastic. It adds or removes cache regions when client requirements and memory availability changes. It also efficiently migrates cache regions when a remote memory region becomes unavailable.
- Implementation and evaluation with a production key-value store. We deploy Redy with FASTER [39] to improve its performance when the hot set is larger than local memory. We measure its improvement using the YCSB benchmark.

The paper is organized as follows. Section 2 shows that data centers have a lot of unallocated memory that could be used as remote caches and configuring RDMA to use it is challenging. Section 3 describes Redy's architecture. Sections 4 and 5 explain how Redy optimizes RDMA for a given workload and SLO. Section 6 discusses VM allocation and cache migration in response to memory changes. Our experimental evaluation of Redy is in Section 7. Section 8 describes the integration of Redy with FASTER. Section 9 covers related work. Section 10 concludes with future work.

## 2 MOTIVATION
## 2.1 Underutilized Cloud Memory

We take it is as given that stateful applications would benefit from more memory. There is a lot of it in data centers, waiting to be utilized. All major data center operators and cloud providers report that memory is highly underutilized. Studies of traces from Google [23, 54], Microsoft [18, 42], Alibaba [9, 26], and Facebook [25] report memory utilization is under 50% and has strong temporal volatility. We confirm these results for unused memory, and extend them by analyzing the dynamics of stranded memory.

**Unallocated memory.** We define *unallocated memory* as the fraction of DRAM not allocated to any VM or container. We measured unallocated memory in 100 Azure Compute clusters over 75 days. Compute clusters host mainstream internal and external VM workloads and represent the majority of servers compared to storage or other specialized clusters. We selected clusters with at least 70% of CPU cores in use. Each cluster trace contains time, duration, resource demands, and server-ids for millions of VMs. We find strong diurnal patterns; the typical peak-to-trough ratio is 2. At the median (across clusters and time), 46% of memory is unallocated. The tenth and first percentile are 37% and 28%, respectively.

**Stranded memory.** A subset of unallocated memory is stranded. At the median, 8% of memory is stranded. This grows as more VMs and containers are allocated with more than 16% stranded at the 90-th percentile and 23% stranded at the 99-th percentile.

We analyze the amount of stranded memory reachable via RDMA by measuring the number of network switches between a server and stranded memory, shown in Figure 1 as a CDF. Half of all servers can reach 1 TB of memory by traversing just one switch, 30 TB by traversing three switches, and 100 TB by traversing five switches. *A small fraction of servers can even reach 1 PB.* Our analysis shows that stranded memory in a public cloud is too significant to ignore.

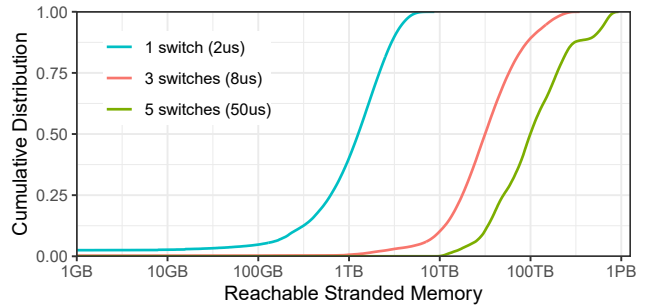

**Figure 1: The significance of stranded memory.**

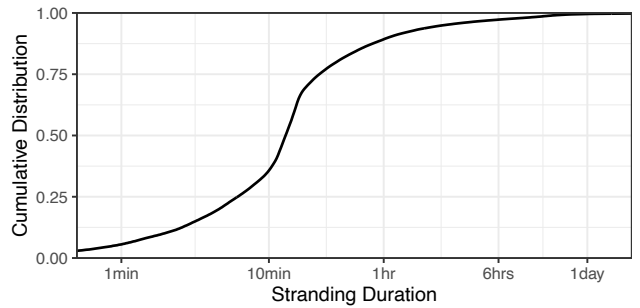

**Figure 2: The dynamics of stranding events.**

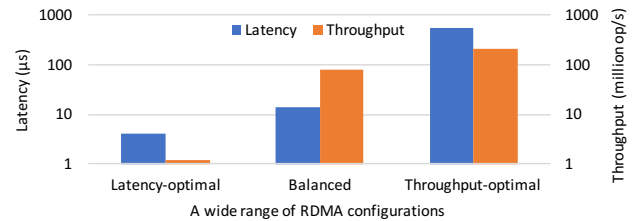

**Figure 3: The impact of the RDMA configuration in Redy.**

**Stranding Duration**. Figure 2 shows the distribution of the duration of stranded memory events. A stranding event begins when a server allocates all CPU cores while ≥1 GB of memory remains unallocated. It ends when a VM or container on the server terminates, making at least one core available. We find that memory is frequently stranded and unstranded with variable durations of minutes to hours. The median stranding event is 13 minutes, with a 25-th percentile of 6 minutes and a 75-th percentile of 22 minutes. Our analysis shows that the amount and duration of stranded memory are highly dynamic, making it challenging to use it effectively.

## 2.2 Diverse RDMA Configurations

We propose using this unallocated memory for RDMA-accessible remote caches. However, optimizing RDMA's performance is hard. Parallelization, asynchrony, thread contention, batching, one-sided vs. two-sided operations, and CPU bottlenecks all affect RDMA throughput and latency. Performance is also highly sensitive to the underlying hardware. Overall, it is difficult to develop a robust solution for a variety of workloads and configurations.

For example, Figure 3 shows the latency and throughput of our caching system, Redy, when writing 8-byte payloads (as in YCSB [17]) to remote memory with three different RDMA configurations. The latency-optimal configuration has 4.1$\mu$s latency, which includes 2.9 $\mu$s network latency, but the throughput is only 1.2 million operations per second (MOPS). The throughput-optimal configuration achieves 205 MOPS, but the latency is 538 $\mu$s. The balanced configuration is in between with 14 $\mu$s latency and 77 MOPS. We have similar findings for reads and other record sizes.

Many configuration parameters affect throughput, latency, and cost. They often improve one performance metric and degrade another. For example, increasing the number of operations in each RDMA transfer (called the *batch size*) increases throughput but also increases latency per operation. Increasing the number of in-flight transfers improves utilization of an RDMA connection and hence its throughput, but it increases latency. Increasing the number of hardware threads that service RDMA requests on the client and server increases throughput, but also increases cost. These conflicting trade-offs imply the need for optimization.

To solve this optimization problem, we need a software architecture that can dynamically tune these parameters, and an optimization algorithm that finds the optimal point in the parameter space. To address these challenges, we propose that the user guides the choice of configuration by specifying an SLO consisting of the desired throughput and latency of the remote cache. It is the system's job to choose the lowest-cost RDMA configuration that satisfies the SLO and then deploy it. Relating cache performance to application performance is out of scope and a possible topic of future work.

## 3 REDY ARCHITECTURE

### 3.1 Design Principles

Redy is a cache service that offers underutilized cloud resources to memory-intensive applications. Its design goals are:

(1) *Generality and ease of use.* Redy must have a flexible interface that can be easily integrated with a variety of memory-intensive cloud applications.
(2) *Customizable performance.* Cloud applications have diverse throughput and latency requirements. Users can customize Redy's performance by providing SLOs for I/O throughput and latency and trade performance for lower cost.
(3) *High resource utilization and minimal disruption.* Redy can exploit underutilized resources and stranded memory, thereby improving cloud resource utilization. This utilization improvement should not disrupt existing applications.
(4) *Robustness to dynamics.* Resource utilization changes over time. One server may become busy while another becomes underutilized. Redy handles such dynamics, offering robust service as long as resources are accessible somewhere.

### 3.2 Back End

Figure 4 shows the architecture of Redy. The front end is implemented by the *Redy client*, which is colocated with its application. It talks to its cluster's back end, which consists of a global cache manager and a set of cache servers that run as VMs. We describe the back end in this subsection and the front end in the next one.

Redy's *cache manager* interacts with the cluster's *VM allocator*. It tracks the available server resources, which it uses to provision VMs. The cache manager offers three operations for allocating a cache: *Allocate*, to allocate one or more VMs for a cache; *Reallocate*, to revise a cache allocation; and *Deallocate*, to drop a cache.

The *Allocate* operation takes three parameters: the desired *amount of memory*, an *SLO* that specifies the desired latency and throughput of reads and writes, and a *duration* that specifies the likely lifetime of the cache. The SLO supports the second design goal by enabling the application to customize the cache's performance, for example, by specifying low latency for an interactive application that requires fast response time or high throughput for an analytics application that does data ingestion and query processing. A *duration* of infinity says that the caller is willing to pay full price for a cache that remains active until it is explicitly deallocated or fails. Shorter durations are meant to benefit from spot pricing of excess resources that the cloud vendor is unable to sell at full price [10, 24, 43], thereby improving resource utilization, the third design goal.

To process an *Allocate* request, the cache manager allocates one or more VMs, each of which consists of memory and zero or more cores, and derives an RDMA configuration that will support the requested SLO. It then returns a list of the allocated VMs and the RDMA configuration to use to communicate with them.

If the cache manager cannot satisfy the requested combination of capacity, SLO, and duration, then the *Allocate* request fails. The request has no effect and the cache manager returns an exception to the client. If a VM is a *spot instance*, then the VM allocator is free to *reclaim* the VM's resources, e.g., to sell the resources for a higher price. In this case, the VM allocator alerts the cache manager of the change and gives it time to compensate for the loss of resources. Today's cloud providers give an early warning of 30-120 seconds.

When the cache manager is notified that a VM failed or was reclaimed, it alerts the Redy client, which must be able to cope with the loss. Ideally, it can provision and populate a replacement VM. This reconfiguration activity is a key challenge for Redy. Its solution addresses the fourth design goal. Details are in Section 6.

The *Reallocate* operation is used to reconfigure an existing cache. The data in the cache can be truncated or remain unchanged depending on the parameters in the reallocation. The *Deallocate* operation is called to release all VM resources for a cache.

Each VM that hosts a cache runs a *cache server*, which is an agent that processes *Connect*, *Read*, and *Write* operations. These operations depend on RDMA details and are described in Section 4.

### 3.3 Front End

A cache client provides a *virtual storage device* abstraction that supports a contiguous byte-addressable address space. The client maps that address space to *memory regions* of the cache's VMs. The size of a memory region is configurable (1 GB by default). The application can perform a read or write operation on the device at an arbitrary address and of arbitrary size (bounded by cache capacity). The client translates the operation into a read or write at the corresponding offset of a memory region. This general device abstraction supports the first design goal.

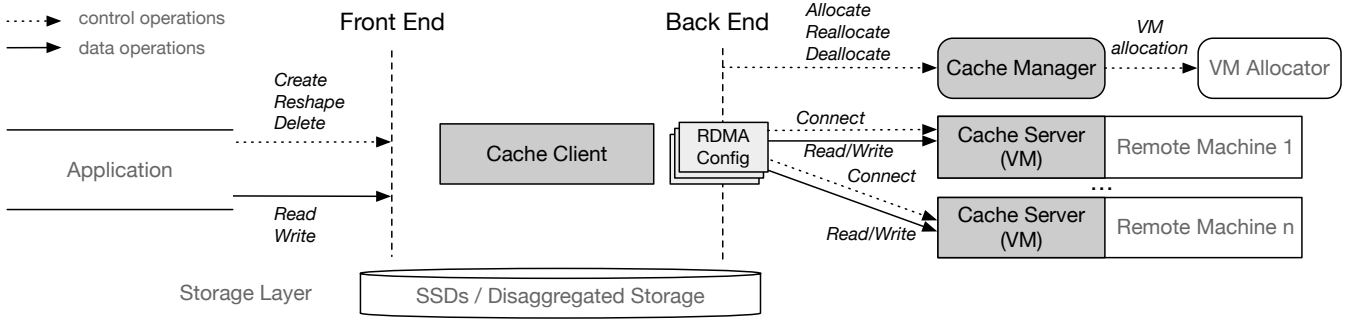

**Figure 4: Architecture of Redy.** An application interacts with the *cache client*. The global *cache manager* asks the data center's VM allocator to reserve VMs to host the cache. *Cache servers* on those VMs coordinate with the client for cache accesses.

| API | Function |
| --- | --- |
| *Create(capacity, SLO, duration [, file])* | Create a cache with the specified capacity, performance SLO, and duration. Optionally populate the cache with a prefix of the file length 'capacity', and return the ID of the created cache. |
| *Read(ID, dst, addr, size, cb)* | Read (async) from a cache with specified address, size, and the callback. |
| *Write(ID, src, addr, size, cb)* | Write (async) to a cache with specified address, size, and the callback. |
| *Reshape(ID, capacity, SLO)* | Change the configuration of a cache with new capacity and SLO. |
| *Delete(ID)* | Delete a cache with specified ID. |

**Table 1: APIs provided by the cache client for applications. Underlined are functions for performing I/Os.**

Table 1 lists the client's APIs to create, manage, and access a cache. The *Create* function creates a cache of a given size, performance level SLO, and duration, and optionally initializes its content based on a file. The SLO specifies a maximum average latency and minimum average throughput of reads and of writes. If *Create* can allocate the requested capacity and the cache can satisfy the SLO and duration, then the client receives a list of VMs and the RDMA configuration for the cache and populates it (if the *file* parameter is present); otherwise, it has no effect and returns an exception.

On receiving the list of VMs, the client constructs a *region table* that maps the cache's address space [0, *capacity*) to memory regions on servers. It divides the address space into *virtual regions*, mapping each one to a *physical region* on a VM (see Figure 5). To service a Read or Write for cache address *x*, the client uses the region table to translate *x* into the address on the VM where *x* is stored.

The two data access operations, *Read* and *Write*, are asynchronous, which is important for performance as we explain in Section 4. When an I/O operation finishes, its associated callback is invoked.

The *Reshape* function enables an application to change the SLO or capacity of a given cache. There are two cases: the SLO changes or it is unchanged. In the first case, the client calls *Allocate* to find new VMs of the requested size that satisfy the SLO. If it succeeds, the client migrates the old cache to the new one, truncating the end of the cache if it shrank. Then it deallocates the old cache. If *Allocate* fails, the cache is unchanged and the client returns an exception.

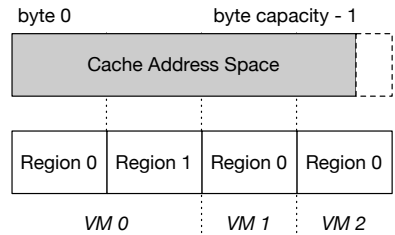

**Figure 5: A region table maps a cache to VMs.**

In the second case, the client resizes the cache. If the cache shrank, the client truncates it. If that frees up regions, the client calls *Reallocate* to notify the cache manager. If the cache grew, the client extends the address space. If the last region has insufficient unused space, then the client calls *Reallocate* to request more VMs.

If the client succeeds in reshaping the cache, it updates the region table. If it cannot allocate enough memory or cannot satisfy the SLO, then it returns an exception and the cache remains unchanged.

The *Delete* function removes a cache by sending *Deallocate* to the manager. Any later access to the cache will return an exception.

## 4 REMOTE CACHE WITH RDMA

This section presents the internals of a Redy cache, specifically, how it configures RDMA to access remote memory regions. The next section describes how Redy provides customized cache performance.

### 4.1 Cloud RDMA Background

RDMA enables an application on a VM to send requests to its NIC to read or write memory on another VM. It uses *kernel bypass*, which means the application interacts directly with its VM's NIC. The transfers are handled entirely by the NICs, with no OS involvement.

An application talks to its NIC via one or more *queue pairs* (QPs), each of which consists of two workqueues: a *send queue* and a *receive queue* for submitting and receiving requests respectively. Each workqueue has an associated completion queue. Multiple workqueues may share the same completion queue.

Communication can be one-sided or two-sided. With one-sided RDMA, the client application directly accesses the server's memory via *read* and *write* operations. A read operation includes the address and length of the server data to be read and the client location where the data should land. Conversely, a write operation includes the address and length of the client data and the server location where the data should land. The client application polls the completion queue for an event that indicates the operation finished.

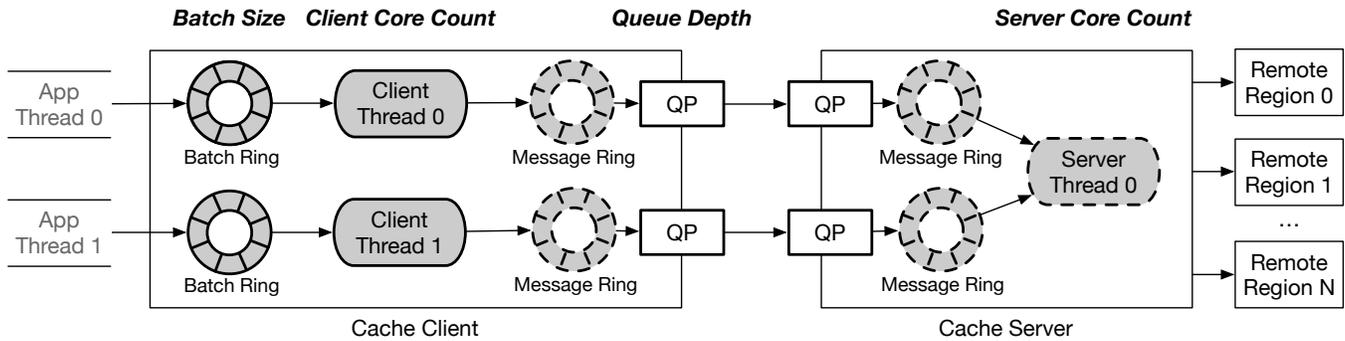

Figure 6: RDMA message flow in Redy. Ring buffers enable pipeline parallelism between adjacent threads.

Every RDMA operation requires memory buffers registered to the NIC, which the NIC can directly access. For example, the RDMA buffer for a *read* operation is used to hold the data that was read. When the read operation is completed, the NIC transfers the data that it read from remote memory into that buffer.

Two-sided RDMA offers *send* and *receive* RPC-like operations in which the server CPU processes the client's request. Redy implements two-sided communications, but like previous work [20, 28, 55, 62] does it using one-sided RDMA *writes*, since they are faster.

One-sided RDMA uses session-oriented communication. A QP can only communicate with the QP that it connects to. Messages are delivered in order with no loss or duplicates.

### 4.2 Cache Implementation

**Connection Setup.** To process *Create* and *Reshape* operations or replace a failed/reclaimed VM, the client asks the cache manager to allocate new VMs. The allocate operation returns a list of VMs and the RDMA configuration to the client. After the client updates the region table, it builds *RDMA connections* by sending a *Connect* message to the cache server on each newly allocated VM. The message includes the number of physical regions the cache uses on the VM and the RDMA configuration. The latter specifies how the client and server communicate: whether communications is one-sided or two-sided, and if two-sided, then how many server CPU cores the cache can use to process RDMA requests. The server allocates the requested number of memory regions, registers them to the NIC, and replies with RDMA access-tokens, one per region, that the client uses to access server memory. When the client receives replies for all *Connect* messages, the cache is ready to use.

**Reads and Writes.** Redy implements reads and writes on a cache as remote memory accesses (see Figure 6). The *Read* and *Write* APIs are asynchronous, so an application can issue requests without waiting for previous ones to finish. The Redy client is multithreaded. Each thread collects read and write requests from an application thread in a *request batch* data structure, which it sends to the server using RDMA. The batch size is configurable from one to hundreds.

Each *server thread* polls messages from one or more RDMA connections. Upon receiving a request batch, it executes the requests on local memory regions. For a write request, the server thread copies the request's payload to the destination address. For a read request, it copies the requested data from the requested memory address to the response buffer. Finally, it sends a *response batch* that contains the results of all requests to the client through the same RDMA connection on which it received the request batch.

Each client thread polls its RDMA connection to retrieve response batches. For each read response in a batch, the client thread copies the payload to the application buffer specified by the corresponding read request. The client invokes the callback function of each read and write request to complete it.

Redy guarantees that all asynchronous requests are executed in order: requests from an application thread are batched in program order, batches are delivered in order with reliable RDMA connections, and they are processed in order by server threads.

### 4.3 Static Optimizations

Redy's RDMA architecture is optimized to exploit RDMA characteristics. Figures 7 and 8 show the effectiveness of each optimization. Unless otherwise mentioned, latency is the time in $\mu$s to process one I/O, which is a Redy read or write call, and throughput is the rate in MOPS. Figure 7 shows the median network round trip latency (light blue), and the median (dark blue) and 99-percentile tail (line with a top) of overall latency. This test uses one application thread, one client thread, and one server thread to read and write 8-byte records in a 1 GB cache with a batch size of one. (Section 7 describes the setup and presents more results.) The details are as follows.

**Lock-free Communications.** To minimize the overhead of exchanging data between threads, we use lock-free ring buffers. Specifically, a client thread accepts I/O requests from an application thread using a *batch ring buffer*, each element of which is a request batch. When a batch becomes full and the RDMA connection is available for another RDMA operation, the client thread moves the batch to its *message ring buffer*, which is registered to the NIC as RDMA buffers. There is a message ring on the server for every connection. Batch and message rings are based on previous work on lock-free ring buffers using atomic compare-and-swap and fetch-and-add [31] and using ring buffers for RDMA data transfer [20], but are customized for the Redy architecture. These ring buffers allow many requests to be passed and processed efficiently from the client to the server. This optimization eliminates data contention compared to a baseline where application threads use locks to share data with client threads, thereby reducing tail latency by 7× and improving throughput by 68.7%, as shown in Figures 7 and 8.

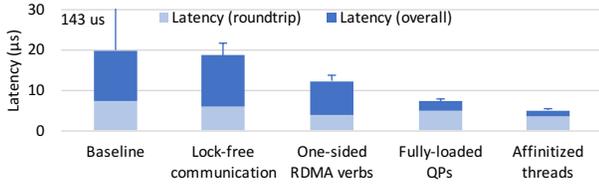

Figure 7: Redy optimizations effectively decrease latency.

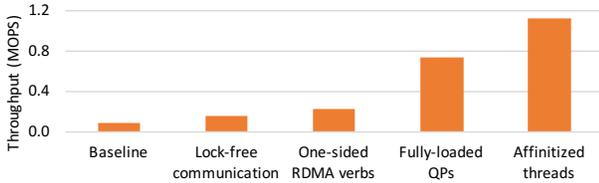

Figure 8: Effectiveness on increasing throughput.

**One-sided Operations.** If a request batch has only one read (or write) request, we translate it to a one-sided *read* (or *write*). Otherwise, we use a *write* to send the request batch to the message ring on the server. This optimization reduces median latency from 19 $\mu s$ to 12 $\mu s$ and increases throughput by 45.3%.

**Fully-loaded Queue Pairs.** The number of in-flight RDMA operations on a connection is called its *queue depth*, which we control by the message ring size. Increasing it reduces waiting time of requests in the batch ring and thus their latency. It also increases network utilization. Compared to one in-flight operation, a queue depth of four reduces latency to 7.1 $\mu s$ and increases throughput from 0.22 MOPS to 0.74 MOPS, a 3.4× speedup. However, the network latency increases with queue depth due to higher traffic, e.g., comparing the light-blue bars for one-sided RDMA and fully-loaded QPs in Figure 7. Although throughput improves when we increase queue depth from four to eight, latency worsens. We measure the performance impact of queue depth, starting from one, and choose the maximum value that improves both latency and throughput.

**NUMA-aware Affinitized Threads.** OS thread scheduling can negatively affect application performance [33, 51]. To avoid this, we pin Redy threads to physical cores in a NUMA-aware fashion. Each client thread is affinitized to an application thread's NUMA node, which reduces communication overhead between threads and stabilizes communication between client threads and the NIC. This achieves a latency of 5 $\mu s$ and throughput of 1.1 MOPS, a 30% and 52% improvement respectively over non-affinitized threads.

## 5 SLO-DRIVEN CONFIGURATION

### 5.1 Performance Variables

RDMA can transfer messages in just a few microseconds. At that time scale, small changes in the instruction count, synchronization delay, memory contention, or processor cache contention can greatly affect RDMA latency and throughput. These effects can be controlled by the choice of RDMA configuration and how it is used. However, since optimal choices depend on the size of cached records and the relative importance of latency and throughput, the choice is necessarily workload dependent.

Based on microbenchmarks and the rich literature on RDMA performance, we have identified four variables that are the primary determinants of Redy cache performance. They are summarized in Table 2. Increasing the value of each variable will increase throughput. But it also increases network traffic, which in turn increases the latency of individual requests. Details are as follows.

| Variable | Description | Lower Bound | Upper Bound |
|---|---|---|---|
| $c$ | the number client threads that process request batches | 1 | client cores |
| $s$ | the number of cache server threads | 0 | $c$ |
| $b$ | the number requests in a batch | 1 | $\lceil \frac{4\,\text{KB}}{\text{record size}} \rceil$ |
| $q$ | the number of in-flight operations | opt. | NIC spec |

Table 2: Variables balancing latency and throughput.

- *Client core count* ($c$) - Increasing client threads adds more computation and RDMA connections for more parallelism. This parameter is capped by available CPU cores in the client VM.
- *Server core count* ($s$) - Increasing threads on the remote server to process batched requests reduces the load on each thread. No server threads are needed if requests are not batched. Each client thread has one RDMA connection, and the server has at most one thread per connection (since the bottleneck of a connection is the network, not server compute), so we cap $s$ at $c$, i.e., $s \leq c$.
- *Batch size* ($b$) - Batching small requests improves network bandwidth utilization. In our RDMA tests, bandwidth utilization and throughput stop improving beyond 4 KB data transfers. Therefore, we cap the batch size at $\lceil \frac{4\,\text{KB}}{\text{record size}} \rceil$ messages.
- *Queue depth* ($q$) - Based on the fully-loaded QP optimization, additional in-flight operations improve bandwidth utilization, i.e., increasing throughput but also latency, similarly to $b$. The upper bound is specified by the NIC, which is 16 in our testbed on Azure HPC clusters [41].

In addition to the trade-off between latency and throughput, there is a trade-off between performance and cost: increasing $c$ and $s$ increases the client and server VM cost.

### 5.2 SLO-based Search

A major challenge of Redy's design is to find an RDMA configuration that satisfies each cache's SLO. Our solution is a two-phase search algorithm: (1) offline modeling and (2) online searching. In offline modeling, we perform measurements to build a function that captures the effect of the configuration parameters ($c$, $s$, $b$, $q$) on latency and throughput. In online searching, we use the function to search for values of these variables that satisfy the latency and throughput specified by the SLO. Our detailed design is below.

**Configuration Space.** An *RDMA configuration* is a tuple [$c$, $s$ $b$, $q$] of configuration parameters. Our performance model is a function $f$ that maps each RDMA configuration to I/O latency and throughput:

$$f : (c, s, b, q) \rightarrow (\text{latency}, \text{throughput})$$

Given the highest number of client cores $C$, the largest batch size $B$ defined by the record size, and the NIC-specific queue depth $Q$, the total number of configurations can be calculated as

$$(\sum_{c=1}^{C}(c+1)) \times B \times (Q - opt.) - C \times (B-1) \times (Q - opt.)$$

where we consider several configuration constraints: (1) the server core count is from zero and to the client core count; (2) if there are no server threads, then batching is disabled so the batch size is one; (3) the minimum queue depth is optimized by the fully-loaded QP technique. Overall the configuration space is $O(C^2 \times B \times Q)$.

In both modeling and searching, we explore the configuration space by incrementally increasing the value of every parameter in a resource-efficient fashion to minimize cost: explore the configurations that do not increase the hardware cost, i.e., increasing $b$ and $q$, before the configurations that do, i.e., $c$ and $s$. We increase $c$ before $s$ to minimize use of limited compute resources on remote memory servers, e.g., in a memory-disaggregated environment.

Formally, we define a Redy configuration space as a *five-level tree*. The root represents configuration options for $s$, the second level for $c$, the third for $b$, the fourth for $q$, and the leaves for latency and throughput. An internal node and the edges below it represents a parameter and its values in increasing order from left to right. A root-to-leaf path represents a configuration. A leaf is the latency and throughput of the path's configuration.

The construction of the tree enforces the aforementioned constraints. For example, all $b$ nodes have only one child ($b = 1$) in the sub-tree of $s = 0$, and the $c$ node under $s = S'$ has $C - S' + 1$ children (from $S'$ to $C$). To explore the space, we do a pre-order traversal to visit configurations that require fewer server and client threads, and thus reduce overall hardware cost.

**Offline Modeling**. We use offline measurements to build a performance model (the function $f$). The model is sensitive to network latency, which varies depending the network distance between the cache client and cache server (cf. Figure 1). We build a performance model for each distance in a data-center-scale deployment. A typical data center network has three distances: one switch (intra-rack), three switches (intra-cluster), and five switches (inter-cluster).

The built-in measurement application on the client VM (the largest VM type for the deployment) allocates a server VM with enough cores (also for the largest configuration of interest). It then creates a Redy cache with an arbitrary configuration. The client starts the modeling by telling the manager the number of available cores for the cache, the record size, and the NIC-specific queue depth. The manager builds the tree representing the configuration space, with empty leaves. The manager and the client then repeatedly generate the next configuration to measure (❶) (see Figure 9), switch to that configuration, measure its latency and throughput by performing I/O operations on the cache, and report the result to the manager (❷). When the manager determines that the model is complete (❸), it signals the application to terminate.

**The Challenge and Solution.** The performance modeling is done offline, when Redy is deployed in a new cloud RDMA environment.

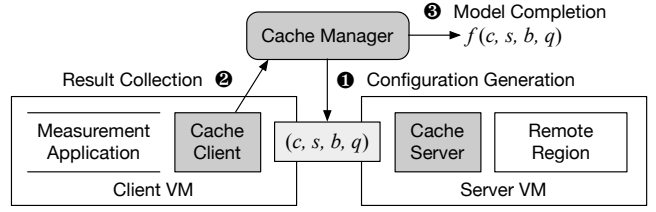

Figure 9: Configuration performance modeling.

Still, the size of the configuration space poses a challenge. In our testbed, a VM has up to 60 cores, of which we assume half are available to a Redy cache, and the NIC-specific queue depth is 16. The model for 8-byte records has ~3M configurations per network distance. If one measurement takes a minute, including switching to the new configuration, performing I/Os, and reporting the result, then building the model takes over five years to finish! So we cannot measure every configuration.

Our solution applies *interpolation* and *early termination*. With *interpolation*, we only measure configurations where parameter values are powers of 2, and we assume a linear growth of latency and throughput between adjacent measured configurations. For example, $f(1, 1, 1, 3)$ is estimated as the mean of $f(1, 1, 1, 2)$ and $f(1, 1, 1, 4)$. This effectively reduces the number of measurements to $O((\log C)^2 \times \log B \times \log Q)$, which is less than two thousand configurations in the above example.

*Early termination* removes unnecessary measurements. Ideally, increasing the value of each variable increases the throughput. However, due to factors such as thread and connection contention, increasing a parameter might not improve throughput while increasing latency. When this happens, we stop measuring configurations where only the value of that particular parameter increases. For instance, if the throughput does not improve from $f(4, 2, 2, 2)$ to $f(8, 2, 2, 2)$, there is no point in measuring $f(16, 2, 2, 2)$.

These two optimizations reduce the number of measurements for the above example to 1000, which took only 15 hours. Section 7 shows the accuracy of the estimated performance by interpolation. The resulting model will remain accurate if the hardware is stable, i.e., the NICs and switches. When hardware changes, the model should be updated by repeating the modeling, but we speculate that such hardware changes are infrequent, once every few years.

**Online Searching**. When the cache manager receives an *Allocate* request, it searches for a configuration to satisfy the given SLO. It uses the algorithm sketched in Figure 10, which traverses the configuration tree in pre-order with pruning to speed up the process.

Line 1 finds the model for the record size specified in the SLO. Line 2 allocates an empty configuration, which is used as the current configuration during the search. Line 3 invokes the traversal function, starting with the root of the model, $s$. If the traversal succeeds, then the algorithm returns config, *which is guaranteed to have the fewest server threads among all possible configurations and thus incurs minimal cost*; otherwise, it returns *null* (Lines 4-6).

If the current visited node is a leaf (Line 8) and the current configuration violates the latency SLO, then the traversal function returns

```
 1  model ← find the model for the record size
 2  config ← empty configuration
 3  result ← Traverse(model.root, SLO, config, 1)
 4  if result = SUCCESS then
 5      return config
 6  return null

 7  Function Traverse(node, SLO, config, level):
 8      if level = 5 then
 9          if node.latency > SLO.latency then
10              return INVALID
11          if node.throughput ≥ SLO.throughput then
12              return SUCCESS
13          return CONTINUE
14      p ← the parameter at this level
15      node_result ← INVALID
16      foreach child in node's children from left to right do
17          config.p ← edge value to child
18          child_result ← Traverse(child, SLO, config, level+1)
19          if child_result = SUCCESS then
20              return SUCCESS
21          if child_result = INVALID then
                  //pruning remaining children
22              return node_result
23          if child_result = CONTINUE then
24              node_result ← CONTINUE
25      end
26      return node_result
```

**Figure 10: Online SLO-based searching in the manager.**

an "invalid" status (Lines 9-10). If latency and throughput are satisfied, then the search returns "success" (Lines 11-12). Otherwise, the traversal explores internal nodes (Line 13). Line 14 identifies the parameter for the current level, and Line 15 initializes the search result as "invalid". Then Lines 16-25 visit the children of the current node left-to-right. For each child, it updates the current configuration parameter with the edge value and then recursively traverses the subtree rooted at the child (Lines 17-18). If the traversal succeeds, the search stops (Lines 19-20). If it returns "invalid", we can safely prune all the remaining children; since increasing the parameter value can only increase the latency, the latency SLO is violated for all of them (Lines 21-22). Finally, if the current child returns "continue", then the next child is visited (Lines 23-24).

In a test to search 100 random SLOs in a space of three million configurations, pruning reduces the number of explored leaf nodes by 25%. The average search time was only 0.027 seconds. Section 7 shows the quality of the returned configurations.

## 6 REMOTE MEMORY MANAGEMENT

### 6.1 Resource Allocation

Recall from Section 3 that an application invokes *Create* to provision a cache of a given capacity, SLO, and duration. The cache client services the function by issuing an *Allocate* to the cache manager.

First, the cache manager translates the capacity and SLO into an RDMA configuration for each network distance, as described in Section 5.2. Then it allocates a VM whose memory and CPU cores are sufficient for the RDMA configuration. Since there are different RDMA configurations for different distances, the cache manager has to find the best VM for the configuration associated with each network distance and then choose the least expensive one.

The cache manager must choose VMs from the menu of VM sizes offered by the cloud provider. Each VM size has fixed cores and memory. Today, providers offer relatively few VM sizes with a high ratio of memory to cores and no VMs consisting of stranded memory. A wider range of choices would enable the manager to choose VMs that more closely match the desired RDMA configuration.

Since the set of VM types changes infrequently, the cache manager can maintain a static list of VM types, with each one's memory size and core count, and its price in each cloud region. To service an allocation request, it identifies the VM types in the client's data center with enough memory and cores and chooses one that has lowest cost and is available within the required network distance.

Beyond these static allocation strategies, there are many ways the cache manager can optimize the choice of VMs. They depend on the optimality criteria it uses and on the VM allocation mechanism of the cluster computing platform it runs on. In some cases, it may be cheaper for the cache manager to select two or more VMs that together satisfy the configuration. Each VM's core-to-memory ratio must be at least that of the configuration, to satisfy the SLO.

Additional cost savings are possible with a spot VM. This is an attractive choice if the cache can be migrated within the 30-120 seconds notice before its VM is reclaimed. This constraint argues for the use of many small VMs instead of a large one, to leave time to migrate each VM cache. We describe migration shortly.

Recent research has shown how to predict the lifetime of spot VMs [11]. This would enable the allocation of VMs that satisfy the requested duration. It could also suggest preemptively migrating a VM's cache, knowing it will likely be reclaimed soon.

At any given time, different VM types might have spot instances available. The cache manager can exploit such cost-saving opportunities by periodically issuing an allocation request for a cheap VM and migrating the cache to it when it becomes available.

The ability of the cache manager to optimize the choice of VMs could be improved by enriching the VM allocator's API. For example, to avoid having the cache manager poll for cheap VMs, the VM allocator could offer an option to alert the cache manager when spot VMs of a certain type become available. It could also offer an option to request the cheaper of one large VM or several smaller VMs based on current spot pricing. VM allocation for spot instances is an active research area. We discuss some recent work in Section 9.

### 6.2 Dynamic Memory Management

If the VM hosting a cache fails or is reclaimed, then the cache client is notified and must allocate another cache to replace it. For a failure, the cache client can use a copy of the cache to populate the new cache. For a reclamation, the cache client can migrate the cache's content to a new cache. The affected parts of cache are unavailable during recovery and possibly during migration. Afterwards, the entire cache is available and must satisfy its performance SLO.

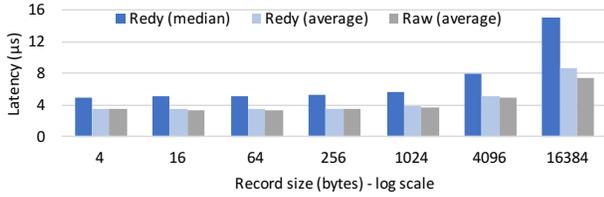

(a) Read latency with record sizes from 4 B to 16 KB.

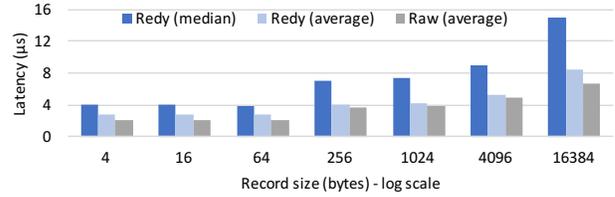

(b) Write latency with record sizes from 4 B to 16 KB.

Figure 11: The latency of Redy caches with latency-optimal configurations for different record sizes.

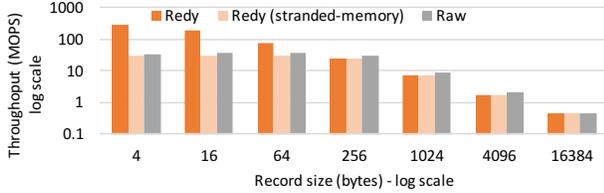

(a) Read throughput with record sizes from 4 B to 16 KB.

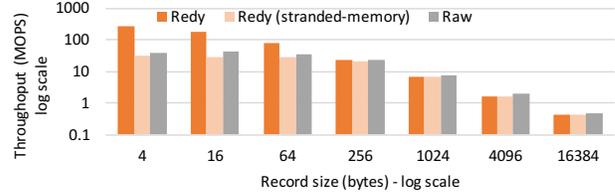

(b) Write throughput with record sizes from 4 B to 16 KB.

Figure 12: The throughput of Redy caches with throughput-optimal and stranded-memory configurations.

The migration period depends in part on the time to provision a new VM. This might exceed the minimum time delay before the spot VM is reclaimed. If this risk is unacceptable or if a VM failure is too disruptive, the cache manager could hold pre-provisioned VMs as targets for migration. Another alternative is replicating the cache. Replica synchronization techniques can be found in [30, 52].

The migration speed also depends on the transfer rate. A tuned RDMA transfer in Redy can fully utilize the network bandwidth.

**Migrating a Cache.** To migrate the content of an existing cache to a newly allocated VM, the cache client needs to tell the new VM to establish a bandwidth-optimized connection with the existing cache. The new VM uses one-sided reads to copy data from the old VM. During the migration, operations on the migrated regions should be paused until the migration is finished. To minimize this performance impact, we employ two optimizations for reads and writes respectively: *unpaused reads* and *pause-on-migration writes*. In *unpaused reads*, we use the old VM to service read operations, and immediately switch to the new VM when the migration is over.

Unlike reads, writes have to be paused during the migration. But instead of pausing all writes (and dependent reads) to the cache, in *pause-on-migration writes*, we migrate regions one by one and pause writes only to the region being migrated. After a region has been migrated, the cache client updates its region table using the new VM and resumes paused writes. When all regions are migrated, the client signals the old VM to terminate. Section 7 evaluates the impact of migration on read and write performance, with and without the optimizations.

**Resizing a Cache.** In response to a *Reshape* invocation, the cache client executes operations to grow or shrink the size of the cache. To grow a cache, the client first uses any memory headroom available in the cache's last VM. For additional growth, the client allocates another VM, using the same memory-to-core ratio, batch size, and queue depth as existing VMs. Depending on the price of spot VMs, it could be cheaper (although more disruptive) to allocate a larger VM and migrate the content of the old VM to the new one. After the new VM is allocated, the client updates its region table. The cache client stalls I/O operations while the cache is being resized.

## 7 EVALUATION

### 7.1 Methodology

**Implementation.** The implementation of Redy consists of 13700 lines of C++ code. It includes the cache client library for applications, cache manager, cache server (shown in Figure 4), and measurement application (in Figure 9). The client library has a Common Language Runtime (CLR) wrapper covering all APIs in Table 1, to enable access by applications in other languages, such as C#.

RDMA transfer in Redy uses the native RDMA library in Windows, NDSPI [40], which supports all RDMA operations. NDSPI has been used to implement other RDMA-based systems, e.g., FaRM [20]. We implemented an RPC framework based on RDMA for efficient operations between clients, servers, and the manager.

**Testbed Setup.** We evaluate Redy on a Microsoft Azure High Performance Computing cluster [41] using the Standard_HB60rs VMs. Each VM has 60 vCPUs based on two 2.0 GHz AMD EPYC 7551 processors, 228 GB of memory, and a 700 GB Azure premium SSD. We run Windows Server 2019 Datacenter as the OS. Each VM is RDMA-enabled using an NVIDIA Mellanox ConnectX-5 NIC [7].

### 7.2 Overall Cache Performance

We first show the overall performance of Redy caches. In this evaluation, we vary data size from very small records (4 bytes) to large blocks (16 KB). For each size, we set up a cache with latency-optimal and throughput-optimal configurations. The purpose of this evaluation is to show Redy's optimal performance for each metric and for different sizes. We compare Redy's cache performance with the raw RDMA network. We measure the latter using the official benchmark tools from Mellanox [6], i.e., nd_read_lat and nd_write_lat for latency, and nd_read_bw and nd_write_bw for throughput.

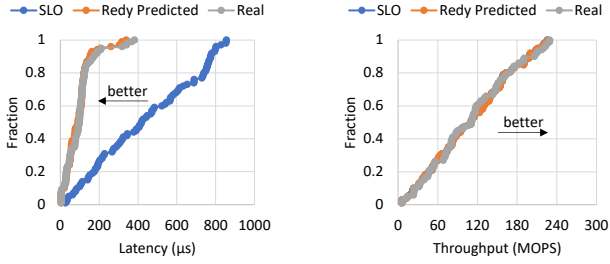

Figure 13: The accuracy of satisfying latency SLOs.

Figure 14: The accuracy of satisfying throughput SLOs.

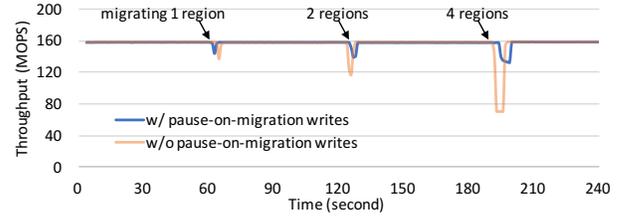

Figure 16: The impact of region migration on writes.

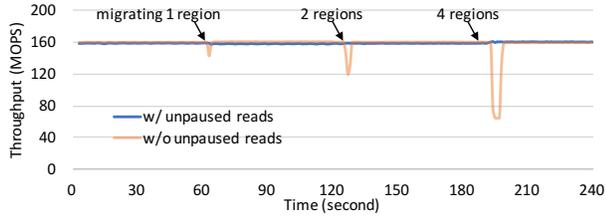

Figure 15: The impact of region migration on reads.

Figures 11a and 11b show the results of latency benchmarking for reads and writes, respectively. Average latency is close to that of the raw network, 3-4 $\mu s$, showing the effectiveness of Redy's latency optimizations described in Section 4.3. An interesting finding is that the write latency is significantly lower than the read latency for records smaller than 256 bytes. This is because a small amount of data to be written can be *inlined* as a parameter in the RDMA *write* invocation, thereby avoiding the latency of fetching the data from main memory to the NIC through the PCIe buses. Inlining no longer works when the data exceeds a threshold (172 bytes in our testbed), so the latency increases. In general, the latency is steadily low until 4 KB records and increases significantly after that.

Figure 12 shows the results for throughput. Read and write throughput are similar. For example, both reading and writing 16 bytes can achieve about 200 MOPS, an order of magnitude higher than raw network throughput, showing that Redy batching is effective at utilizing the bandwidth. When the record size increases, throughput drops as fewer operations/second are needed to saturate the network. But up to 256 bytes, Redy performs much better than the raw network.

All latency-optimal configurations use one-sided memory access using no server cores, so Redy is particularly cheap for this case. Conversely, for record sizes up to 1KB, high-throughput configurations work best if they have a few cores to support batching.

Between latency-optimal and throughput-optimal configurations, there is a big space of configurations that make trade-offs between latency and throughput. We let the applications customize cache performance using their SLOs.

### 7.3 Performance Customizability

Offline modeling builds interpolated performance models whose accuracy determines whether they can satisfy users' SLOs. The speed of online searching determines how fast we can configure a cache. To evaluate both, we measure the accuracy of the model for the three million configurations in Section 5.2 and the time to search configurations for given SLOs using the algorithm in Figure 10.

We draw 100 performance SLOs between the lowest and highest latency and throughput values in the model. An SLO consists of cache latency and throughput, which are drawn independently from a uniform distribution. For each SLO, we search the configuration space for one that Redy predicts will satisfy the SLO. We then configure the cache based on this configuration, measure its latency and throughput, and compare them with the SLO. The accuracy of the model is defined by how close the predicted latency and throughput mimic the real ones. High accuracy means that the real performance will satisfy the SLO.

Figures 13 and 14 show the results for both latency and throughput. Each figure shows three CDFs—the SLO, predicted, and real performance—of the corresponding metric. Since we draw SLOs randomly between the lowest and highest values, the SLOs in both figures are spread uniformly across their ranges. Figure 13 shows that the predicted and real latency are close: 92 $\mu s$ vs. 98 $\mu s$ at the median, and 206 $\mu s$ vs. 212 $\mu s$ at the 95th percentile. They are all lower than the requested latency—satisfying the SLO. Figure 14 shows findings for throughput: the predicted and real throughput values are 110.5 MOPS and 110.7 MOPS respectively at the median, both closely matching the requested throughput of 110.4 MOPS, and are 211.5 MOPS and 219.3 MOPS at the 95th percentile, also close to the requested 211.4 MOPS. The latency of the caches is much lower than the SLOs, while the throughput just reaches the SLOs, because the searching algorithm in Figure 10 starts from low-latency low-throughput configurations and gradually moves toward high-latency and high-throughput ones. This matches our cost-efficient goal: the average client and server core counts of the resulting configurations are 7.3 and 1.5.

Redy is also fast at finding configurations. The time spent on searching for the right configuration for an SLO in the space is 2 $\mu s$ to 0.12 s with an average of 0.027 s and a median of 0.01 s, achieving interactive speed for cache allocation.

### 7.4 Robustness to Dynamics

When a VM that hosts a part of a cache is to be reclaimed, the cache client requests a new VM (or multiple VMs) from the manager and migrates the affected regions using a throughput-optimized configuration. In our testbed, it takes 1.09 s to online migrate a 1 GB region. This argues for using spot VMs of ≤ 27 GB, to ensure they can be migrated within 30 s. Thus, it is feasible to use spot VMs that are available for only a short time, say a few minutes.

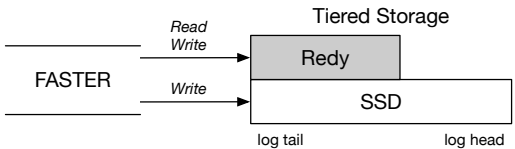

**Figure 17: FASTER with Redy.** New records are appended to both tiers. Reads to records in Redy are only served by Redy.

We evaluate the performance impact of region migration using a cache that consists of seven 1 GB regions. Initially, all regions are hosted in one VM. We run this cache with 8-byte records for four minutes and migrate one, two, and four regions at the second, third, and fourth minute respectively to a different VM. We measure the throughput change. Figures 15 and 16 show that the throughput of both reads and writes drops by around 15%, 25%, and 57% in the migration of one, two, and four regions without optimizations. By contrast, the read throughput with *unpaused reads* is unaffected by the migration, and the write throughput with *pause-on-migration writes* decreases by at most 15%, no matter how many regions are migrated. This demonstrates that Redy minimizes the impact of resource dynamics on its caches.

## 8 FASTER WITH REDY

FASTER is a high-performance open-source key-value store that is used at Microsoft and elsewhere [15, 39]. It is an example of a stateful cloud service that can benefit from using a remote cache, as discussed in Section 1.1. We integrate Redy with FASTER to demonstrate its ease of use and practical value.

### 8.1 Data Organization in FASTER

FASTER runs as a multi-threaded library in the address space of an application client. It has a hash index that maps keys to record addresses. The index is stored in the client's memory.

FASTER stores records in a *hybrid log* where the tail of the log is stored in main memory and the remainder is spilled to storage, such as a server-attached SSD or a cloud storage service. The log is organized as a sequence of *segments*. The tail of the main-memory section supports in-place updates. The rest is read-only.

A read operation looks up a record in the index and then retrieves it from memory or storage. To insert a record, it is appended to the tail and added to the index. To update a record in the read-only portion of the log, it is appended to the tail in main memory and its index entry is updated. To free up main memory, the oldest segment of the read-only main memory section of the log is appended to storage. To free up storage, the oldest segment is read, its reachable records are appended to the log tail, and then it is deallocated.

### 8.2 Integrating Redy

FASTER clients access storage through an interface called `IDevice`, which exposes storage as a byte-addressable sequential address space. FASTER supports *tiered storage*, which is a "meta-device" that wraps a set of `IDevice` implementations, called *tiers*. Each tier is smaller and faster than the next higher tier, and is a replica of a suffix (i.e., tail) of the higher tiers [32]. FASTER services a read operation from the lowest tier that has the data.

To keep the tiers consistent, an append operation is applied to all tiers. It is acknowledged to the client after all tiers have applied the append. A user can alter this semantics via FASTER's *commit point* setting, which is the lowest tier whose commit denotes the completion of an update. This is useful for committing quicker than the highest tier, which may be very slow.

We integrate Redy as an `IDevice` in this tiered storage, as the first tier (see Figure 17). An SSD is the second tier, which contains the entire log. Thus, reads are serviced by Redy if the record is stored in the Redy cache. Otherwise, it is serviced by the SSD. Cloud blob storage could be a third tier, as a highly-available backup.

### 8.3 Evaluation

We evaluate the performance of FASTER with Redy using the YCSB benchmark [17] in the same cloud environment as Section 7. We compare with two alternatives: a device that only uses local SSD; and a device that accesses remote memory using SMB Direct, an RDMA-enabled file server protocol with higher throughput and lower latency than the regular Windows file server [1]. Throughput is the critical metric for this benchmark, so we configure the Redy cache for high throughput. Our YCSB database contains 250 million key-value records (8-byte key and 8-byte value), ~6 GB in total in FASTER. Every operation is a read governed by either a uniform distribution or a Zipfian distribution ($\theta = 0.99$). Additionally, we use a value size of 1 KB, resulting in a ~260 GB database.

Figure 18a shows the throughput of FASTER in MOPS on the uniform workload, with different storage devices. In this experiment, we give FASTER 1 GB of local memory, and the remainder of the log is spilled to the device. In the tiered device, we allocate an 8 GB Redy cache so that all operations are served by Redy. When there is one thread, FASTER achieves 0.8 MOPS with Redy while SMB Direct and SSD are much lower with less than 0.1 MOPS. With two threads, the throughput with Redy increases to 1.6 MOPS. With SMB Direct and SSD it improves to 0.15 MOPS, but that still is 10× lower than Redy. Adding more threads improves FASTER's performance with all devices, but the gap between Redy and other alternatives remains large. Figure 18b shows the results with the Zipf distribution where data accesses are skewed. FASTER uses local memory to cache frequently-accessed records, which reduces load to the devices. Hence, the throughput is higher than that with the uniform distribution for all devices. However, when we decrease the available local memory for caching in FASTER (similarly when we increase the database size), both the absolute throughput and the relative difference between Redy and other devices become closer to that of the uniform distribution, as shown in Figure 18c.

FASTER with Redy achieves higher throughput for large records as well. Figure 18d shows that with four threads, the throughput of accessing records with 1 KB values is 0.9 MOPS with Redy, 8× and 20× higher than with SMB Direct and SSD respectively. Figures 18e–18h show that even when the client has a local cache as large as 10 GB, 20 GB, 40 GB, and 80 GB respectively, the tail of the Zipfian distribution still bottlenecks the overall performance.

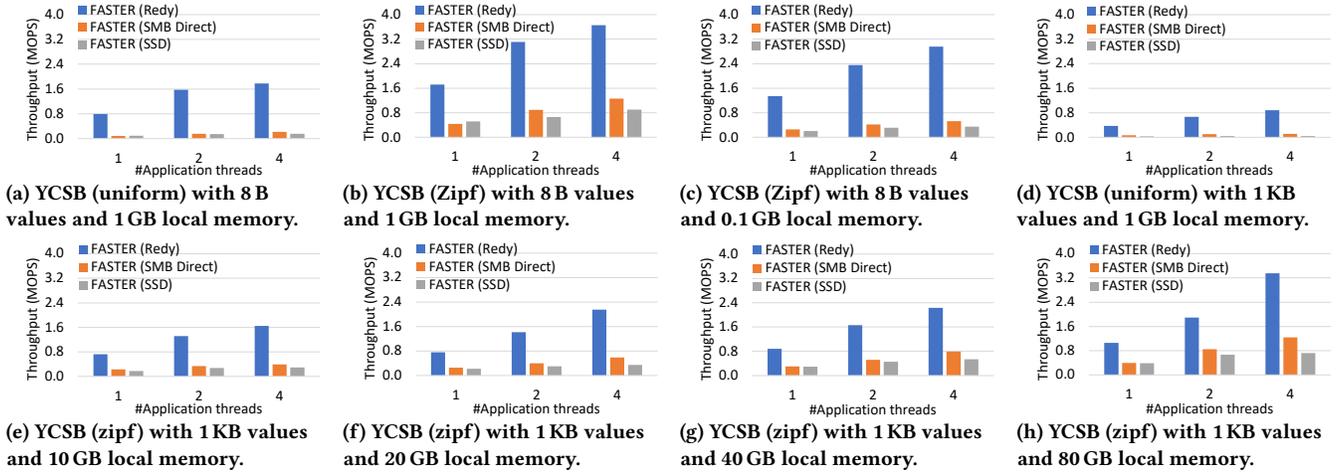

(a) YCSB (uniform) with 8 B values and 1 GB local memory.

(b) YCSB (Zipf) with 8 B values and 1 GB local memory.

(c) YCSB (Zipf) with 8 B values and 0.1 GB local memory.

(d) YCSB (uniform) with 1 KB values and 1 GB local memory.

(e) YCSB (zipf) with 1 KB values and 10 GB local memory.

(f) YCSB (zipf) with 1 KB values and 20 GB local memory.

(g) YCSB (zipf) with 1 KB values and 40 GB local memory.

(h) YCSB (zipf) with 1 KB values and 80 GB local memory.

Figure 18: The performance with Redy, SMB Direct, and SSD when FASTER's working set is larger than local memory.

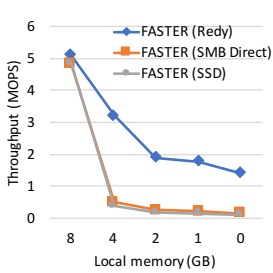

Figure 19: FASTER with various local memory sizes.

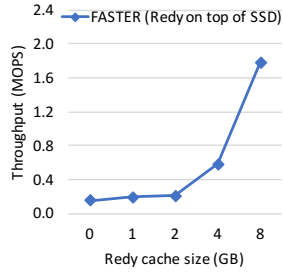

Figure 20: Tiered store with various remote cache sizes.

Spilling requests to Redy has at least 2× higher throughput than other cloud services, i.e., SMB Direct and SSD storage.

Figure 19 varies the size of local memory used by FASTER (with four threads). With 8 GB local memory, FASTER services all (uniform) operations from local memory, achieving high throughput of 5 MOPS. When we spill the entire log to the storage device, FASTER achieves 1.4 MOPS using Redy, vs. 0.15 MOPS and 0.12 MOPS for SMB Direct and SSD. Compared to local memory only, the performance of FASTER with Redy decreases by 72% (vs. 97% with SMB Direct and 98% with SSD); but it saves memory cost by 100%, since it uses stranded memory, which is essentially free.

To show the impact of the cache size in the tiered device we vary the Redy cache size from 0 to 8 GB with 1 GB client local memory (Figure 20). As expected, performance increases significantly when more cache is allocated.

In summary, when FASTER's working set exceeds local memory, spilling data to a Redy cache results in better performance than spilling to the RDMA baseline or SSD. We note FASTER using synchronous local-memory outperforms the asynchronous device interface due to I/O code path and context switching overheads. As new high-throughput devices such as Redy become commonplace, we believe this is an important area for future optimization.

## 9 RELATED WORK

Redy is an RDMA-accessible remote dynamic cache targeted for data centers. No systems that we know of offer Redy's SLO-based configuration and dynamic reconfiguration. We summarize related systems and explain the differences as follows.

**Cache Servers.** A cache server is an in-memory distributed key-value store that supports access by a large number of clients. It is typically used to store content that is accessed over the Internet. Popular cache servers are Memcached [2] and Redis [3]. By contrast, Redy offers inexpensive remote caches in the cloud environment. CompuCache [58] is a cloud service that supports both data caching and compute offloading. However, since it uses RPC, it cannot use stranded memory.

**Disaggregated Memory.** The systems community has been exploring the use of disaggregated memory for over a decade. Lim et al. [35, 36] propose the use of specialized memory servers. The hypervisor extends the memory of a VM by servicing page faults from memory servers. Infiniswap [25] avoids specialized hardware by using remote memory as a paging device accessed via one-sided RDMA. Aguilera et al. [8] exposes remote memory as files. LegoOS [47] is a disaggregated OS that emulates Linux APIs with RDMA implementations. Gao et al. [22] investigate network requirements to support disaggregated memory without degrading application performance. Zhang et al. [59, 60] describe the benefit and performance overhead of using disaggregated memory for DBMSs. In contrast to the above work, Redy abstracts remotely accessible memory as a cache, rather than as process memory or as a file.

**RDMA.** RDMA has been a subject for research in the database, systems, and networking communities for many years [21]. Herd [28] and FaRM [20] are RDMA-accessible key-value stores. FaRM also supports multistep transactions, as does [12, 13]. DFI [53] provides a data flow abstraction based on RDMA. Cai et al. [14] propose a distributed shared memory framework with an RDMA-based memory coherence protocol. Liu et al. [37] optimize the bandwidth of RDMA specifically for shuffles. Ziegler et al. [62] report on microbenchmarks of RDMA. Li et al. [34] explore RDMA performance benefits to a DBMS via SMB Direct. Redy is different from these works in its cache design that supports fine-grained data accesses and performance customizability. Kalia et al. [29] provide RDMA developers with guidelines for low-level RDMA optimizations. In comparison, Redy hides RDMA complexities with an easy-to-use cache API. Kalia et al. also explore the benefit of batching, but speculatively and only for unconnected QPs.

**VM Scheduling and Migration.** Redy's cache manager uses the cluster VM scheduler to allocate VMs for caches. The challenges of allocating VMs for large data centers are discussed in [19, 27, 44, 46]. Redy's allocator is rather unique in requiring a minimum amount of memory that can be partitioned across multiple VMs, each VM satisfying a minimum ratio of cores to memory.

Redy migrates cache when its VM fails or is evicted. This is similar to VM migration, but without the need to freeze program execution to move its state. Some past work on VM migration includes [16, 49, 50, 56]. To mitigate the effect of VM eviction, researchers are exploring dynamic alternatives, where VMs can shrink or grow to offer all unallocated resources on the server where it runs [11, 48]. Extending Redy's ability to exploit dynamic resource allocation is an interesting avenue for future work.

## 10 CONCLUSION AND FUTURE WORK

This paper described Redy, a cloud service that provides high performance caches using RDMA-accessible remote memory. Redy automatically configures resources for a given latency and throughput SLO and automatically recovers from failures and evictions of remote memory regions. We integrated Redy with a production key-value store, FASTER. The experimental evaluation shows that Redy can deliver its promised performance and robustness.

One future challenge is sizing the server-local memory cache. This depends on several factors: the required response-time and throughput, the cache miss rate as a function of cache size, and the latency of servicing a cache miss. The first factor is determined by the application. The second factor depends on the degree of skew of the access distribution, which in turn depends on the application and data layout. The third factor depends on the type of storage that services the miss, such as remote memory, server-local disk, or cloud storage. If there is more than one storage tier, then the average latency is a sum of the latency of each tier weighted by the fraction of misses serviced by that tier.

A second challenge is to integrate the cache manager's cache allocation with the cluster's VM allocator. We expect this will benefit from extensions to the latter, as discussed in Section 6.1.